\documentclass[11pt]{article}

\usepackage[margin=1in]{geometry}
\usepackage[T1]{fontenc}
\usepackage{amsmath,amssymb}
\usepackage{graphicx}
\usepackage[numbers,sort&compress]{natbib}
\usepackage[hidelinks]{hyperref}
\usepackage{url}
\usepackage{microtype}

\title{Stage-Supervised Latent Reasoning for Single-Shot JavaScript Deobfuscation}

\author{
Rong Feng \qquad Suman Saha\\[0.35em]
Dept. of Computer Science and Engineering\\
The Pennsylvania State University\\
\texttt{rjf5768@psu.edu} \qquad \texttt{szs339@psu.edu}
}

\date{}

\begin{document}

\maketitle

\begin{abstract}
JavaScript obfuscation is widely used to protect code, but it also makes program analysis and security review substantially harder. Existing LLM-based deobfuscation methods usually treat the task as one-step translation, ignoring the staged structure of practical deobfuscation pipelines. This WIP paper proposes a stage-aware latent reasoning framework that converts intermediate outputs from a deterministic deobfuscation tool into supervision for Coconut-based training. The model learns from multi-stage rewrites during training but generates the final cleaned program in a single shot at inference time. Preliminary results on JsDeObsBench show that the Coconut-based model improves syntactic validity to 50\%, compared with 15\% for direct fine-tuning and 25\% for a zero-shot baseline, and reaches 80\% semantic correctness among valid outputs, indicating better semantic faithfulness than either comparison model in deobfuscation.
\end{abstract}


\section{Introduction}

Code obfuscation is widely used to protect intellectual property and conceal malicious logic~\cite{schrittwieser2016protecting}. In JavaScript, where code is easily distributed and often exposed on the client side, obfuscation is especially common. Although it can serve legitimate goals, it also makes program comprehension, security auditing, vulnerability analysis, and maintenance substantially harder~\cite{schrittwieser2016protecting}. For analysts and reverse engineers, the core challenge is to recover clean, analyzable code from heavily obfuscated programs. This challenge is especially difficult because real-world obfuscated JavaScript rarely relies on a single transformation. Instead, it often combines string encoding, control-flow flattening, dead-code injection, and self-defending constructs~\cite{laszlo2009obfuscating,schrittwieser2016protecting}. As a result, deobfuscation is not a single rewrite step, but a staged process that progressively restores readable structure.

Traditional analysis techniques provide only partial relief. Static analysis becomes less effective when control flow and data flow are intentionally distorted, and dynamic approaches can be brittle and input-dependent~\cite{salwan2018symbolic,schrittwieser2016protecting}. Large language models (LLMs) have recently shown strong promise for code understanding and transformation tasks~\cite{xie2024resym}, but prior work shows that off-the-shelf LLMs remain unreliable for direct deobfuscation of complex programs, often producing outputs that look plausible yet fail to compile or preserve semantics~\cite{chen2025jsdeobsbench}. A central weakness of current LLM-based deobfuscation is that it is usually framed as a one-step translation problem from obfuscated code to clean code~\cite{chen2025jsdeobsbench}. That formulation overlooks the fact that practical deobfuscation pipelines proceed through multiple intermediate simplification stages. Deterministic tools already expose these intermediate states, but existing LLM-based methods largely ignore them. At the same time, purely deterministic tools remain limited to predefined rewrite patterns and can struggle with unseen obfuscation variants or new combinations of transformations~\cite{salwan2018symbolic,schrittwieser2016protecting}.

In this work, we propose a \textit{stage-aware latent reasoning framework} for JavaScript deobfuscation. Our key idea is to turn intermediate outputs from a deterministic deobfuscation tool into supervision for model training. We generate multi-stage deobfuscation sequences and train a model using the Coconut (Chain of Continuous Thought) paradigm~\cite{hao2024coconut}. Coconut enables reasoning in continuous latent space by feeding the model's hidden state back as input embeddings, allowing it to represent multi-step transformations without explicitly generating intermediate text~\cite{hao2024coconut}. Unlike conventional chain-of-thought prompting, which requires explicit natural-language intermediate steps~\cite{wei2022chain}, Coconut operates entirely in latent space~\cite{hao2024coconut}, making it well suited to program transformation tasks whose intermediate states are structured code artifacts rather than textual explanations.

During training, the model observes the full sequence of deobfuscation stages. During inference, it receives only the obfuscated program and must generate the cleaned code in a single shot. This design combines the structural guidance of deterministic pipelines with the generalization ability of LLMs. We evaluate the approach on JsDeObsBench~\cite{chen2025jsdeobsbench} and study whether stage-aware latent reasoning improves recovery of readable and semantically faithful JavaScript code.

In our preliminary evaluation, the Coconut-based model achieves the strongest overall performance, improving syntactic validity to 50\%, compared with 15\% for direct fine-tuning and 25\% for the zero-shot base model. Among syntactically valid outputs, it also reaches 80\% semantic correctness, compared with 0\% for direct fine-tuning and 40\% for the zero-shot base model. These results suggest that stage-aware latent reasoning improves both syntactic robustness and semantic faithfulness in JavaScript deobfuscation.
\section{Related Work}

Prior work on deobfuscation has largely followed two directions: rule-based program analysis and data-driven neural or language-model methods. Our work sits between them by asking whether deterministic intermediate transformations can serve as structured supervision for latent reasoning, rather than being used as final outputs or discarded entirely.

\subsection{Code Obfuscation and Deobfuscation}

Code obfuscation is widely used to protect intellectual property and conceal malicious logic, but it also makes program analysis and reverse engineering harder~\cite{schrittwieser2016protecting}. Common transformations such as control-flow flattening and opaque predicates distort program structure while preserving behavior~\cite{laszlo2009obfuscating,schrittwieser2016protecting}. Traditional deobfuscation methods rely on symbolic reasoning, pattern matching, and related program analysis techniques~\cite{salwan2018symbolic}. These approaches can be effective for specific obfuscation patterns, but their effectiveness often declines when multiple transformations are composed or when control structure becomes highly irregular~\cite{schrittwieser2016protecting}.

\subsection{Large Language Models for Code Analysis}

Recent language-model-based methods have shown strong promise for code recovery and related understanding tasks, including symbol and variable recovery~\cite{xie2024resym}. However, recent evidence shows that off-the-shelf LLMs remain unreliable for robust JavaScript deobfuscation, often producing outputs that appear plausible but fail to compile or preserve semantics~\cite{chen2025jsdeobsbench}. A central limitation of current LLM-based deobfuscation is that it is usually framed as a direct translation problem from obfuscated code to clean code, without modeling the intermediate transformations that arise in practical deobfuscation pipelines~\cite{chen2025jsdeobsbench}. This leaves an important gap between how real deobfuscation tools operate and how current LLM-based systems are trained. Our method addresses this gap by treating the intermediate states of a deterministic deobfuscation pipeline as structured supervision, rather than using tool outputs only as final outputs or auxiliary artifacts.



\subsection{Chain-of-Thought and Latent Reasoning}

Chain-of-thought prompting improves reasoning by encouraging models to generate intermediate steps in natural language~\cite{wei2022chain}. For program transformation, however, the most useful intermediate states are often structured code artifacts rather than textual explanations. Coconut addresses this issue through continuous latent reasoning, where the model feeds its last hidden state back as the next input embedding instead of decoding intermediate tokens~\cite{hao2024coconut}. While Coconut has shown promise on reasoning tasks~\cite{hao2024coconut}, its use for staged program transformation and deobfuscation has not yet been explored. Our work addresses that gap by combining tool-generated deobfuscation stages with latent reasoning supervision.
\section{Approach}

\begin{figure}[htbp]
\centering
\includegraphics[width=\textwidth]{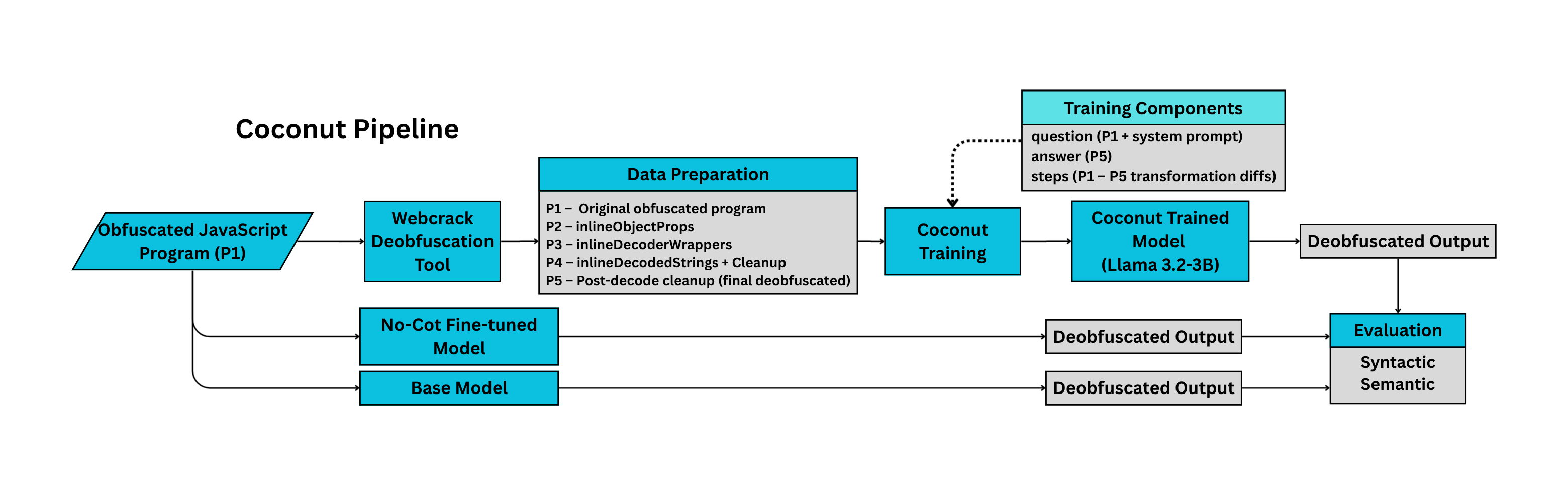}
\caption{Overview of the stage-aware deobfuscation framework. Multi-stage intermediate programs ($P_1$--$P_5$) provide supervision during Coconut training, while inference performs single-shot deobfuscation; outputs are compared with No-CoT and base models using syntactic and semantic evaluation.}
\label{fig:approach}
\end{figure}

Our goal is to train a model that maps obfuscated JavaScript directly to clean, readable code in a single shot. To do this, we build a stage-aware training pipeline around deterministic intermediate rewrites produced by the Webcrack\footnote{\url{https://github.com/j4k0xb/webcrack}} deobfuscation framework.

Figure~\ref{fig:approach} shows the workflow. Given an obfuscated program, Webcrack applies a sequence of normalization passes that progressively simplify the code. We capture these intermediate states and use them to supervise the latent reasoning process of a Coconut-based language model. During training, the model observes the transformation trajectory through structured step-level supervision. Instead of storing full intermediate programs, we encode each transition from $P_i$ to $P_{i+1}$ as a labeled diff. This gives fine-grained supervision at the transformation level while avoiding the redundancy of full program snapshots. At inference time, the model receives only the obfuscated input and must generate the final cleaned program directly. This design combines the structure of deterministic rewriting with the generalization ability of large language models.

\subsection{Dataset and Stage Extraction}

We construct our dataset from 1{,}302 obfuscated JavaScript programs from JsDeObsBench~\cite{chen2025jsdeobsbench}. To fit hardware memory limits, we first retain the 300 smallest programs by file size, then randomly sample 100 for experiments, split into 60 training, 20 validation, and 20 test examples. For each program, we run a modified version of Webcrack to record the intermediate output after each deterministic rewrite pass, yielding the transformation trajectory
\[
P_1 \rightarrow P_2 \rightarrow P_3 \rightarrow P_4 \rightarrow P_5
\]

Each stage corresponds to a Webcrack transformation, or a small group of related transformations, that progressively reduces obfuscation while preserving program behavior. \textit{$P_1$} is the raw obfuscated JavaScript program, typically containing string-array encoding, hexadecimal indexing, opaque boolean expressions, decoder aliases, and array shuffling logic. \textit{$P_2$} removes alias-based indirection introduced through object properties by inlining property accesses that reference decoder functions. \textit{$P_3$} inlines thin wrapper functions around the main string-array decoder so calls resolve directly to the core decoder. \textit{$P_4$} resolves encoded strings to concrete values and removes the associated string array, rotation logic, and decoder definitions. \textit{$P_5$} performs final cleanup, including structural simplifications such as string merging, dead code elimination, and partial control-flow simplification. Together, these stages form a deterministic trajectory that removes obfuscation artifacts and restores readable program structure.

\subsection{Training Formulation and Latent Supervision}

Each training example is represented as a triplet \texttt{(question, steps, answer)} in the Coconut framework~\cite{hao2024coconut}. The \textit{question} field presents the obfuscated program $P_1$ and instructs the model to generate the final deobfuscated program $P_5$ directly. The prompt is intentionally simple: it defines the task, requires semantic preservation, and asks for syntactically valid executable JavaScript, but it does not specify intermediate steps or transformation rules. Those signals are provided separately through the structured \textit{steps} field, which stores the ordered transformation trajectory from $P_1$ to $P_5$.

\begin{itemize}
    \item \textbf{question} contains the deobfuscation prompt together with the obfuscated input program $P_1$;
    \item \textbf{steps} stores the ordered transformation trajectory from $P_1$ to $P_5$;
    \item \textbf{answer} is the final deobfuscated program $P_5$.
\end{itemize}

The \textbf{steps} field provides stage-specific supervision, including the applied transformation and the corresponding code change from $P_i$ to $P_{i+1}$. This formulation gives the model explicit guidance about how the program is simplified across stages while keeping the representation compact and aligned with the final single-shot generation objective.

We train the model with Coconut, which performs reasoning in continuous latent space rather than through explicit intermediate tokens. In our setting, the Webcrack stage sequence serves as supervision for this latent reasoning process, guiding the model to internalize multi-stage deobfuscation. As training progresses, explicit intermediate reasoning is gradually replaced by latent thought tokens, allowing the model to absorb the deobfuscation trajectory into its internal representations while still generating only the final output.

\subsection{Model Variants and Inference}

Our framework is model-agnostic, but in this study we use Llama 3.2-3B as the shared backbone for all experiments. To isolate the effect of stage-aware latent reasoning, we compare three variants: the zero-shot base model, a no-CoT fine-tuned model trained only on input-output pairs, and a Coconut fine-tuned model trained with latent reasoning supervision. Because all three variants use the same 3B backbone, the comparison reflects differences in training objective rather than model scale.

At inference time, all models receive only the obfuscated program $P_1$ and must generate the final normalized program $P_5$ directly. The base and no-CoT models perform standard single-pass generation, while the Coconut model inserts a fixed number of latent thought tokens before decoding the output. This train-test asymmetry allows the model to learn from multi-stage supervision during training while remaining practical at inference time, when intermediate deobfuscation states are unavailable.
\section{Experimental Results}

We conduct a preliminary evaluation to assess whether stage-aware latent reasoning improves JavaScript deobfuscation over direct fine-tuning and a zero-shot baseline. Our experiments focus on two questions: whether the method produces more syntactically valid outputs and whether those outputs more reliably preserve program semantics.

\subsection{Experimental Setup}

All experiments were conducted on the Pennsylvania State University Roar HPC cluster using NVIDIA A100 GPUs (40GB VRAM). Training was implemented in PyTorch with Fully Sharded Data Parallel across 2 GPUs and run in bfloat16 precision. Both fine-tuned models, Coconut and No-CoT, used the same optimization settings: AdamW, learning rate $1 \times 10^{-4}$, weight decay $0.01$, batch size 1, and gradient accumulation over 8 steps. This shared setup ensures that differences in performance are attributable to the training objective rather than hardware or optimization changes.

\subsection{Evaluation Methodology}

We evaluate outputs using two criteria. First, we check syntactic validity; only programs that parse are considered valid. Second, for syntactically valid outputs, we assess semantic correctness by executing the generated program and the ground-truth deobfuscated program $P_5$ on test inputs and comparing their outputs.

Because this WIP study evaluates only 20 programs, semantic correctness is verified through a combination of automated execution-based comparison and manual inspection. A larger-scale study will require a fully automated semantic evaluation pipeline, potentially using differential testing or symbolic execution techniques~\cite{cadar2008klee}.

\subsection{Results and Discussion}

\begin{table}[h]
\centering
\caption{Syntactic and semantic results on the 20-program test set. Semantic evaluation is performed only on syntactically valid outputs.}
\label{tab:results}
\begin{tabular}{lcc}
\hline
Model & Syntax Valid & Semantic Match \\
\hline
Coconut    & 10/20 (50\%) & 8/10 (80\%) \\
No-CoT     & 3/20 (15\%)  & 0/3 (0\%)   \\
Base Model & 5/20 (25\%)  & 2/5 (40\%)  \\
\hline
\end{tabular}
\end{table}

Table~\ref{tab:results} shows that the Coconut model performs best on both evaluation criteria. It achieves 50\% syntactic validity, compared with 15\% for No-CoT and 25\% for the base model, suggesting that stage-aware latent reasoning produces more well-formed outputs. Among syntactically valid outputs, Coconut also achieves the highest semantic correctness at 80\%, compared with 40\% for the base model and 0\% for No-CoT. These results suggest that stage-aware latent reasoning is more effective than both direct fine-tuning and zero-shot generation for JavaScript deobfuscation, improving both syntactic robustness and semantic faithfulness. At the same time, the findings remain preliminary because they are based on a small test set and partially manual semantic evaluation, so larger-scale validation is needed in future work.
\section{Future Work}

Our immediate next step is to strengthen the empirical evaluation of this WIP. We will expand experiments to a broader benchmark with diverse JavaScript obfuscation patterns, report structural and task-level metrics, compare against stronger baselines, including direct one-shot deobfuscation and deterministic tool-only pipelines, and add ablations over the number of training stages. We will also extend semantic validation through syntax checks, execution-based comparison, and differential testing to ensure that improved readability does not come at the cost of changed behavior. In parallel, we will run a lightweight ablation study to isolate the contributions of stage supervision and latent reasoning, including comparisons against input-output fine-tuning and controlled variation in the number of training stages.

Beyond this initial evaluation, we plan to study whether our framework can analyze obfuscated malicious JavaScript. In that context, recovering readable, semantically faithful code aids not only program understanding but also threat analysis, triage, and investigation. Long-term goals like cross-tool generalization and scaling to long inputs remain important, but our immediate focus is stronger evidence of correctness, robustness, and security relevance.
\section{Conclusion}

This WIP paper presents a stage-aware approach to JavaScript deobfuscation that trains a latent-reasoning model with intermediate transformation stages produced by a deterministic deobfuscation pipeline. The central idea is to move beyond one-step deobfuscation by exposing the model to the deobfuscation process structure during training while preserving single-shot generation at inference time. Our preliminary results suggest that this design improves both syntactic validity and semantic faithfulness over direct fine-tuning and a zero-shot baseline, with the Coconut-based model achieving the strongest performance. These findings indicate that stage-level supervision is a promising direction for improving LLM-based deobfuscation. Future work will expand the evaluation with broader benchmark results, stronger semantic validation, and a focused ablation study to isolate the contributions of stage supervision and latent reasoning.

\bibliographystyle{ACM-Reference-Format}
\bibliography{main}

\end{document}